\documentclass[runningheads]{llncs}

\usepackage{hyperref}
\usepackage[T1]{fontenc}
\usepackage[scaled=0.85]{beramono}
\usepackage{graphicx}
\usepackage{xcolor}
\usepackage{listings,listings-rust}
\definecolor{Green}{rgb}{0.0,0.5,0.0}
\definecolor{White}{rgb}{1,1,1}
\definecolor{Blue}{rgb}{0.0,0.0,1}
\definecolor{Red}{rgb}{1,0,0}
\definecolor{Purple}{rgb}{0.4,0,0.6}
\definecolor{Orange}{rgb}{1,0.65,0}
\definecolor{Gray}{rgb}{0.5,0.5,0.5}
\newcommand{\annot}[1]{{\color{Blue}#1}}

\newcommand{\produced}[1]{{\color{Green}#1}}
\newcommand{\consumed}[1]{{\color{Red}{\underline{\annot{#1}}}}}
\newcommand{\ghost}[1]{{\color{Gray}#1}}

\newcommand{\ConsumeChunk}[1]{\color{Orange}{\textsf{ConsumeChunk}(#1)}}

\DeclareMathAlphabet{\mathsfsl}{OT1}{cmss}{m}{sl}

\newcommand{\nodei}{\mathsfsl{node1}}
\newcommand{\nodeii}{\mathsfsl{node2}}
\newcommand{\nodeiii}{\mathsfsl{node3}}

\newcommand{\reversed}{\mathsfsl{reversed}}
\newcommand{\original}{\mathsfsl{original}}
\newcommand{\next}{\mathsfsl{next}}
\newcommand{\result}{\mathsfsl{result}}

\newcommand{\purl}[1]{\textsf{\url{#1}}}

\begin{document}
\title{Foundational VeriFast: Pragmatic Certification of Verification Tool Results through Hinted Mirroring}
\titlerunning{Foundational VeriFast}
%
\author{Bart Jacobs\orcidID{0000-0002-3605-249X}}
\authorrunning{B. Jacobs}
%
\institute{KU Leuven, Department of Computer Science, DistriNet Research Group\\
\email{bart.jacobs@kuleuven.be}\\
\purl{https://distrinet.cs.kuleuven.be/people/BartJacobs}}
\maketitle              
\begin{abstract}
VeriFast is a leading tool for the modular formal verification of correctness properties of single-threaded and multi-threaded C and Rust programs. It verifies a program by symbolically executing each function in isolation, exploiting user-annotated preconditions, postconditions, and loop invariants written in a form of separation logic, and using a separation logic-based symbolic representation of memory. However, the tool itself, written in roughly 30K lines of OCaml code, has not been formally verified. Therefore, bugs in the tool could cause it to falsely report the correctness of the input program. We here report on an early result extending VeriFast to emit, upon successful verification of a Rust program, a Rocq proof script that proves correctness of the program with respect to a Rocq-encoded axiomatic semantics of Rust. This significantly enhances VeriFast's applicability in safety-critical domains. We apply hinted mirroring: we record key information from VeriFast's symbolic execution run, and use it to direct a replay of the run in Rocq. 

\end{abstract}
\section{Introduction}

This paper is structured as follows: we recall VeriFast \cite{fvf} in \S1, describe the proof script generation in \S2, and discuss soundness in \S3. We finish by discussing limitations and related work, and offering a conclusion in \S4.

\section{Background: VeriFast}

\begin{figure}
\begin{lstlisting}
/*@
pred llist(head: *mut u8) =
    if head == 0 {
        true
    } else {
        *(head as *mut *mut u8) |-> ?next &*& llist(next)
    };
@*/

unsafe fn reverse_iter(original: *mut u8, reversed: *mut u8) -> *mut u8
//@ req llist(original) &*& llist(reversed);
//@ ens llist(result);
{
    //@ open llist(original);
    if original.is_null() { return reversed; }
    let next = *(original as *mut *mut u8);
    *(original as *mut *mut u8) = reversed;
    //@ close llist(original);
    reverse_iter(next, original)
}

unsafe fn reverse(mut list: *mut u8) -> *mut u8
//@ req llist(list);
//@ ens llist(result);
{
    //@ close llist(0);
    reverse_iter(list, std::ptr::null_mut())
}

fn main()
//@ req true;
//@ ens true;
{
    unsafe {
        //@ close llist(0);
        let mut node1: *mut u8 = std::ptr::null_mut();
        //@ close llist(&node1 as *mut u8);
        let mut node2 = &raw mut node1 as *mut u8;
        //@ close llist(&node2 as *mut u8);
        let mut node3 = &raw mut node2 as *mut u8;
        //@ close llist(&node3 as *mut u8);
        let reversed = reverse(&raw mut node3 as *mut u8);
        reverse(reversed);
        std::process::abort();
    }
}
\end{lstlisting}
\caption{An example VeriFast-annotated Rust program that reverses a linked list of pointer variables}\label{fig:reverse}
\end{figure}

Consider the Rust program in Fig.~\ref{fig:reverse}. Function \lstinline|reverse|
takes a pointer to the head of a linked list of pointer variables, reverses the
linked list in-place, and returns a pointer to the head of the resulting linked list (which is the last node of the original linked list). The function is implemented using recursive helper function \lstinline|reverse_iter|, which repeatedly pops a node from the front of the first given linked list and pushes it onto the front of the second one, until the first linked list is empty, and then returns the second one. The main function allocates three pointer variables on the stack, initializes them so that they constitute a linked list with \lstinline|node3| at the head, calls \lstinline|reverse| on it, and then calls \lstinline|reverse| again on the resulting linked list.

This program uses \lstinline|unsafe| blocks; therefore, the Rust compiler cannot guarantee the absence of undefined behavior. In particular, it cannot guarantee that the pointer dereferences are safe. Instead, we apply VeriFast\footnote{\purl{https://github.com/verifast/verifast}} \cite{fvf} to verify the absence of undefined behavior. VeriFast requires that we annotate each function with a precondition and a postcondition, expressed in a variant of separation logic \cite{seplogic-csl01,seplogic-lics02}, using a recursive separation logic predicate \lstinline|llist| to assert ownership of a linked list of pointer variables. A \emph{points-to assertion} \lstinline{*P |-> V} asserts ownership of the variable pointed to by pointer \lstinline|P|, and that it currently stores value \lstinline|V|. The \emph{separating conjunction} \lstinline|A1 &*& A2| asserts that \lstinline|A1| and \lstinline|A2| both hold, \emph{separately}; the conjuncts must assert ownership of \emph{disjoint} sets of variables. The occurrence of \lstinline|&*&| in the predicate definition asserts that the linked list is acyclic; the occurrence in the precondition of \lstinline|reverse_iter| asserts that the two linked lists do not overlap. The right-hand side of a points-to assertion can be a \emph{pattern}; the pattern \lstinline|?X| binds the pointed-to value to logical variable \lstinline|X|, whose scope extends to the end of the enclosing assertion.

\begin{figure}
$$\begin{array}{l}
\annot{[], [], \emptyset}\\
\textrm{\lstinline{//@ close llist(0);}}\\
\annot{[\produced{\texttt{llist}(\mathsf{nullptr})}], [], \emptyset}\\
\textrm{\lstinline{let mut node1: *mut u8 = std::ptr::null_mut();}}\\
\annot{[\consumed{\produced{\mathsf{points\_to}(\nodei, \mathsf{nullptr})}}, \consumed{\texttt{llist}(\mathsf{nullptr})}], [\texttt{\&node1} \mapsto \nodei], \{\nodei \neq \mathsf{nullptr}\}}\\
\textrm{\lstinline{//@ close llist(&node1 as *mut u8);}}\quad\ConsumeChunk{0}; \ConsumeChunk{0}\\
\annot{[\produced{\texttt{llist}(\nodei)}], [\dots], \{\dots\}}\\
\textrm{\lstinline{let mut node2 = &raw mut node1 as *mut u8;}}\\
\annot{[\consumed{\produced{\mathsf{points\_to}(\nodeii, \nodei)}}, \consumed{\texttt{llist}(\nodei)}], [\texttt{\&node2} \mapsto \nodeii, \dots], \{\nodeii \neq \mathsf{nullptr}, \dots\}}\\
\textrm{\lstinline{//@ close llist(&node2 as *mut u8);}}\quad\ConsumeChunk{0}; \ConsumeChunk{0}\\
\annot{[\produced{\texttt{llist}(\nodeii)}], [\dots], \{\dots\}}\\
\textrm{\lstinline{let mut node3 = &raw mut node2 as *mut u8;}}\\
\annot{[\consumed{\produced{\mathsf{points\_to}(\nodeiii, \nodeii)}}, \consumed{\texttt{llist}(\nodeii)}], [\texttt{\&node3} \mapsto \nodeiii, \dots], \{\nodeiii \neq \mathsf{nullptr}, \dots\}}\\
\textrm{\lstinline{//@ close llist(&node3 as *mut u8);}}\quad\ConsumeChunk{0}; \ConsumeChunk{0}\\
\annot{[\consumed{\produced{\texttt{llist}(\nodeiii)}}], [\dots], \{\dots\}}\\
\textrm{\lstinline{let reversed = reverse(&raw mut node3 as *mut u8);}}\quad\ConsumeChunk{0}\\
\annot{[\consumed{\produced{\texttt{llist}(\reversed)}}], [\texttt{reversed}\mapsto\reversed, \dots], \{\dots\}}\\
\textrm{\lstinline{reverse(reversed);}}\quad\ConsumeChunk{0}\\
\\
\ghost{\textrm{Produce precondition \lstinline{llist(original) &*& llist(reversed)}}}\\
\annot{[\produced{\texttt{llist}(\reversed)}, \consumed{\produced{\texttt{llist}(\original)}}], [\texttt{reversed}\mapsto\reversed, \texttt{original}\mapsto\original], \emptyset}\\
\textrm{\lstinline{//@ open llist(original);}}\ \;\raisebox{-.2em}[0pt][0pt]{\reflectbox{\includegraphics[height=1em]{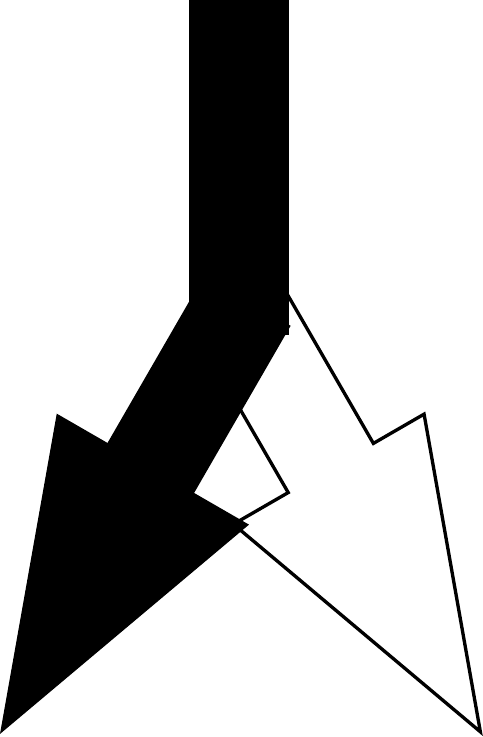}}}\quad\ConsumeChunk{1}\\
\annot{[\produced{\texttt{llist}(\next)}, \consumed{\produced{\mathsf{points\_to}(\original, \next)}}, \texttt{llist}(\reversed)], [\dots], \{\original \neq \mathsf{nullptr}\}}\\
\textrm{\lstinline|if original.is_null() \{ return reversed; \}|}\\
\textrm{\lstinline{let next = *(original as *mut *mut u8);}}\quad\ConsumeChunk{1}\\
\annot{[\consumed{\produced{\mathsf{points\_to}(\original, \next)}}, \texttt{llist}(\next), \texttt{llist}(\reversed)], [\texttt{next}\mapsto\next], \{\dots\}}\\
\ghost{\textrm{Auto-open $\mathsf{points\_to}(\original, \next)$}}\quad\ConsumeChunk{0}\\
\annot{[\consumed{\produced{\mathsf{points\_to\_}(\original, \mathsf{some}(\next))}}, \texttt{llist}(\next), \texttt{llist}(\reversed)], [\dots], \{\dots\}}\\
\textrm{\lstinline{*(original as *mut *mut u8) = reversed;}}\quad\ConsumeChunk{0}\\
\annot{[\consumed{\produced{\mathsf{points\_to}(\original, \reversed)}}, \texttt{llist}(\next), \consumed{\texttt{llist}(\reversed)}], [\dots], \{\dots\}}\\
\textrm{\lstinline{//@ close llist(original);}}\quad\ConsumeChunk{0}; \ConsumeChunk{1}\\
\annot{[\consumed{\produced{\texttt{llist}(\original)}}, \consumed{\texttt{llist}(\next)}], [\dots], \{\dots\}}\\
\textrm{\lstinline{reverse_iter(next, original)}}\quad\ConsumeChunk{1}; \ConsumeChunk{0}\\
\annot{[\consumed{\produced{\texttt{llist}(\result)}}], [\dots], \{\dots\}}\\
\ghost{\textrm{Consume postcondition \lstinline{llist(result)}}}\quad\ConsumeChunk{0}
\end{array}$$
\caption{Symbolic execution traces for functions \texttt{main} and \texttt{reverse\_iter} ($\original \neq \mathsf{nullptr}$ case), showing the symbolic heap, symbolic store, and path condition at each point. Since the path condition only grows, and in these examples so does the symbolic store, we only show new elements. Producing a chunk adds it to the front of the symbolic heap. Newly produced heap chunks are shown in green; heap chunks that will be consumed next are underlined in red. $\mathsf{ConsumeChunk}$ hints are shown in orange (see \S 3).}\label{fig:traces}
\end{figure}

To verify a program, VeriFast symbolically executes each function separately, representing memory as a list of so-called \emph{heap chunks}\footnote{This is a bit of a misnomer. A better name would have been ``chunk of resource'', since heap chunks are also used to represent stack-allocated memory and other resources.}. A heap chunk is either a \emph{points-to chunk} or a \emph{predicate chunk}. A points-to chunk is either of the form $\mathsf{points\_to\_}(p, s)$, asserting ownership of a possibly-uninitialized variable pointed to by pointer $p$, with initialization state $s$, or of the form $\mathsf{points\_to}(p, v)$, asserting ownership of an initialized variable at $p$, with value $v$. Here, $p$, $s$, and $v$ are \emph{terms} of first-order logic, which may contain \emph{symbols} (hence \emph{symbolic execution}). Declaring a local variable whose address is taken \emph{produces} (i.e., adds to the symbolic heap) a $\mathsf{points\_to\_}$ chunk; assigning to a local variable whose address is taken \emph{consumes} (i.e., removes from the symbolic heap, failing if not found) an appropriate $\mathsf{points\_to\_}$ chunk and produces a $\mathsf{points\_to}$ chunk. Calling a function consumes the callee's precondition and then produces its postcondition. Consuming assertion \lstinline|llist(list)| fails unless a $\texttt{llist}(\ell)$ chunk is present in the symbolic heap, where $\ell$ is a term that is provably equal to the value of variable \lstinline|list|. To introduce such predicate chunks, the user must generally insert \lstinline|close p(args);| ghost commands into the program, which consume the body of predicate \lstinline|p| and produce a predicate chunk $\texttt{p}(\emph{args})$. This process can be reversed using \lstinline|open p(args);| ghost commands. Symbolic execution traces for \texttt{main} and \texttt{reverse\_iter} are shown in Fig.~\ref{fig:traces}.

\section{Proof script generation}

VeriFast now optionally emits, upon successful verification of a Rust program, a proof script\footnote{The proof script generated for the example from Fig.~\ref{fig:reverse} can be viewed at \textsf{\purl{https://github.com/verifast/verifast/blob/master/tests/rust/purely\_unsafe/trivialest\_reverse\_expected.v}}} for the Rocq \cite{rocq} system for developing machine-checked mathematical proofs, defining constants \texttt{bodies} (a list of all functions of the program, with their bodies expressed in a slight variant\footnote{See \purl{https://github.com/verifast/verifast/blob/master/bin/rust/rocq/VfMir.v}} of Rust's Mid-level Intermediate Representation (MIR)), \texttt{preds} (a list of all VeriFast predicate definitions), \texttt{specs} (a list with each function's precondition and postcondition), and \texttt{symex\_trees} (a list with for each function, a \emph{symbolic execution tree} containing hints for the Rocq-encoded symbolic execution algorithm), and a single goal \texttt{bodies\_are\_correct preds specs symex\_trees bodies}, expressing the correctness of each function with respect to its specification, with a fixed proof. At least for the example program of Fig.~\ref{fig:reverse}, the proof goes through, implying that the program is indeed correct, assuming that \texttt{bodies} faithfully represents the program, that \texttt{specs} faithfully expresses the desired correctness properties, and that \texttt{bodies\_are\_correct} is sound (discussed in the next section).

\texttt{bodies\_are\_correct}\footnote{Defined in \purl{https://github.com/verifast/verifast/blob/master/bin/rust/rocq/SymbolicExecution.v}, which refers to \purl{https://github.com/verifast/verifast/blob/master/bin/rust/rocq/Values.v} for the definition of type \texttt{Value}.} is essentially a (for now, very incomplete) re-implemen\-tation of VeriFast's symbolic execution algorithm in Rocq. The main differences are as follows:
\begin{itemize}
\item VeriFast sometimes performs \emph{automation steps}. For example, while trying to consume a $\mathsf{points\_to\_}(\ell, \_)$ chunk, if no matching chunk is found but a $\mathsf{points\_to}(\ell, v)$ chunk is found, the latter is \emph{auto-opened} to obtain $\mathsf{points\_to\_}$\-$(\ell, \mathsf{some}(v))$. To simplify the implementation of \texttt{bodies\_are\_correct}, these auto-steps are explicitly specified in \texttt{symex\_trees}.
\item When trying to consume a chunk, e.g.~$\mathsf{points\_to}(\ell, \_)$, if VeriFast finds a chunk $\mathsf{points\_to}(\ell', v)$ in the symbolic heap, it asks an SMT solver whether $\ell = \ell'$. Only if the solver returns \textsf{Valid} does VeriFast consider the chunk to match. \texttt{bodies\_are\_correct}, instead, receives the index $k$ of the chunk that should match as a $\mathsf{ConsumeChunk}(k)$ hint in \texttt{symex\_trees}.
\item VeriFast uses an SMT solver to decide whether the path condition is satisfiable; if not, it abandons the symbolic execution path. \texttt{bodies\_are\_correct} instead receives a $\mathsf{Done}$ hint in \texttt{symex\_trees} when the path should be infeasible. 
\end{itemize}
\texttt{bodies\_are\_correct} (at least, in the current state of the system) reduces to a simple proof goal, intended to be discharged by a simple fixed tactic script that essentially boils down to \texttt{repeat (split || intro || congruence || tauto)}.

\section{Soundness}

We have proven\footnote{See \purl{https://github.com/verifast/verifast/blob/master/bin/rust/rocq/SymexSoundness.v}}, in Rocq, soundness of \texttt{bodies\_are\_correct} (the SymEx) with respect to an axiomatic semantics\footnote{See \purl{https://github.com/verifast/verifast/blob/master/bin/rust/rocq/AxSem.v}} (the AxSem) of VF MIR expressed in Iris \cite{iris}, a framework for separation logic in Rocq. The proof justifies the SymEx's use of VeriFast predicates and function specifications, and its different treatment of locals based on whether their address is taken; none of these appear in the AxSem.
The soundness proof is based on a model\footnote{See \purl{https://github.com/verifast/verifast/blob/master/bin/rust/rocq/LogicalHeaps.v}} of VeriFast symbolic heaps in terms of \emph{logical heaps}, essentially multisets of $\mathsf{points\_to\_}$ chunks (for now). The meaning of VeriFast predicates and assertions is defined inductively in this model, and the soundness of producing and consuming assertions and opening and closing predicates is proven against it.

For now, the axiomatic semantics has not itself been validated in any way. It has one known unsoundness: like VeriFast, for now it ignores \textsf{StorageLive} and \textsf{StorageDead} MIR statements and considers all locals to be alive throughout the body. After fixing this, we would like to verify soundness of the AxSem with respect to one or more other semantics, such as the Radium semantics used by RefinedRust \cite{refined-rust}, or a future Rocq encoding of MiniRust \cite{minirust}.

\section{Conclusion}

In this paper we present a first step towards Foundational VeriFast for Rust, achieved through Hinted Mirroring, a pragmatic approach that is likely to be applicable to other verification tools as well. Much work remains to be done, however, including supporting more Rust features (such as loops, structs, generics, and unwinding) and more VeriFast features (such as inductive datatypes, fixpoint functions, lemmas, fractional permissions, and verifying semantic well-typedness of safe abstractions per RustBelt \cite{rustbelt}), and validating our axiomatic semantics. While our current implementation is minimal, we designed it  with the goal of eventually supporting all of VeriFast, including future VeriFast extensions relating to Rust's pointer aliasing rules and relaxed atomics.

The only existing foundational verification tool for Rust we are aware of is RefinedRust \cite{refined-rust}. Foundational verification tools for other languages include VST \cite{vst,vst-a} and RefinedC \cite{refinedc} for C. All of these, however, are implemented directly inside Rocq, constraining the user experience and preventing the use of SMT solvers for high-performance automation.

Hinted Mirroring is a special case of certificate generation (e.g.~\cite{sledgehammer,smt-coq})%
, and of oracle-guided result validation (e.g.~\cite{compcert}). However, while there exists work towards providing foundational backing for non-foundational program verification tools (such as for Boogie \cite{formal-boogie}, Why3 \cite{why3-coq} and Viper \cite{translational}) we are not aware of existing applications of Hinted Mirroring for separation-logic-based program verification tools.

\begin{credits}

\subsubsection{\discintname}
None.
\end{credits}
%
%
%
\bibliographystyle{splncs04}
\bibliography{foundational-verifast}
\end{document}